\documentclass[a4paper,11pt]{article}
\usepackage{jheppub} 
\usepackage{lineno}
\usepackage{xcolor}
\usepackage{colortbl}
\usepackage[normalem]{ulem}

\usepackage{diagbox}
\usepackage{siunitx} 
\title{\boldmath Normal modes of charged AdS solitons}
\author[a]{Mengqi Lu,}
\author[a,b,c]{Ming Zhang,}
\author[a,b]{Robert B. Mann}

\affiliation[a]{Department of Physics and Astronomy, University of Waterloo,\\
 200 University Ave W, Waterloo, Ontario N2L 3G1, Canada}
\affiliation[b]{Perimeter Institute for Theoretical Physics, \\ 31 Caroline St. N., Waterloo, Ontario N2L 2Y5, Canada}
 \affiliation[c]{Department of Physics, Jiangxi Normal University,\\ Nanchang 330022, China}
\emailAdd{m64lu@uwaterloo.ca}
\emailAdd{mingzhang@jxnu.edu.cn}
\emailAdd{rbmann@uwaterloo.ca}

\abstract{We study massive charged scalar field perturbations in four- and five- dimensional  charged anti-de Sitter soliton spacetimes. Appropriate boundary conditions are established via a local analysis of the perturbation equations. The normal mode spectra are then calculated numerically using the Horowitz-Hubeny method and a collocation method. We reveal  scaling laws and asymptotic behaviors governing the normal mode spectra. The reality of the normal mode frequencies indicates the dynamic stability of the soliton, which in turn provides support for the positive energy conjecture in asymptotically locally anti-de Sitter spacetime.
}

\begin{document}
\maketitle
\flushbottom

\section{Introduction}

According to the correspondence conjecture between anti-de Sitter (AdS) spacetime  and  conformal field theory (CFT) \cite{Maldacena:1997re,Gubser:1998bc,Witten:1998qj,Aharony:1999ti,tHooft:1993dmi,Susskind:1994vu}, the AdS black hole has a holographically dual description in its corresponding CFT. Alternatively, AdS-CFT duality provides  an effective way of studying black hole physics. 
AdS black hole solutions in  gravitational theory can 
have  either spherical, hyperbolic, or planar topology 
of their constant-$r$ sections
(and therefore of the horizon), 
 whose curvatures are positive, negative, and zero, respectively. These topological black holes can be either spherically symmetric \cite{Aminneborg:1996iz,Mann:1996gj,Mann:1997iz,Birmingham:1998nr,Brill:1997mf} or axially symmetric \cite{Klemm:1997ea,Caldarelli:1998hg}. Phase transitions of  spherical black holes \cite{Hawking:1982dh} can be explained by the confining/de-confining phase transition of the large $N$ dual CFT \cite{Witten:1998qj,Witten:1998zw}.  It has been shown \cite{Emparan:1999gf} that the hyperbolic black hole can be described by the thermal Rindler states of the dual CFT. 
 Statistical mechanics of  planar black holes have been  studied via the Dirichlet brane states or the thermal states of $\mathcal{N}=4$ supersymmetric Yang-Mills theory \cite{Klebanov:1996un,Witten:1998zw}. 

The ground state of a planar black hole is known as the AdS soliton \cite{Horowitz:1998ha}, and can be obtained via   double analytic continuation of a planar AdS black hole. The confining ground state \cite{Myers:1999psa} is on a torus with antiperiodic boundary conditions imposed on the fermions  along (at least) one cycle. The cycle of the field theory smoothly caps off in the bulk at a ``bubble'' \cite{Reynolds:2017jfs} or in other words, the infrared floor. As the AdS soliton has a compact dimension, it is only locally asymptotically AdS, and its corresponding CFT on the boundary  has a compact spatial dimension. The AdS soliton has negative energy, which can be identified as the boundary CFT's Casimir energy.

The AdS soliton  has proven to be useful in non-perturbative studies of  strongly coupled Quantum Chromodynamics (QCD)  via the study of solutions to   AdS supergravity within the framework of the AdS/CFT correspondence \cite{Witten:1998zw}. 
To study  QCD in the real world, supersymmetry and conformal invariance 
must each be broken,
the former via  supersymmetry-breaking boundary conditions and 
the latter by
compactification. In supersymmetric field theory, we can compactify one spatial direction on a circle, where antiperiodic boundary conditions are imposed on the fermions. On the AdS side, this scenario can be described by the AdS soliton
\cite{Horowitz:1998ha}, where the supergravity fermions on the circle that  contracts to a point in the interior are antiperiodic \cite{Constable:1999gb}.

In  stability analyses of spacetime, one typically examines both its thermodynamic and dynamical aspects. This approach applies not only to black holes but also to solitons. The thermodynamic stability of the AdS soliton has been investigated via the  phase  analysis, and   Hawking–Page phase transitions \cite{Hawking:1982dh} between the AdS soliton and the planar black holes have been shown to occur  \cite{Surya:2001vj,Banerjee:2007by}. This corresponds to the confinement–deconfinement phase transition in the gauge field.  Phase transitions between charged planar black holes and   planar charged solitons have more recently been found \cite{Anabalon:2022ksf}. The transition temperature is determined by the electric potential of the  black hole and the magnetic flux of the soliton. Recently  it was shown that  a planar AdS  black hole can coexist with two distinct gravitational solitons and there is a new type of  triple point supported by the non-linear electrodynamics \cite{Quijada:2023fkc}. 
The chemistry (and related holographic complexity) of solitons  has also been a subject of recent interest
\cite{Kunduri:2013vka,Andrews:2019hvq,Reynolds:2017jfs,Yang:2023qxx}. 

Compared to the thermodynamic stability analysis, the dynamic stability analysis of the soliton spacetime is less studied. The dynamic stability of an uncharged soliton, obtained from the planar Schwarzschild black hole, has been analyzed in \cite{Csaki:1998qr,deMelloKoch:1998vqw,Zyskin:1998tg,Constable:1999gb}. According to the conjectured duality \cite{Witten:1998zw} between supergravity and 't Hooft large $N$  gauge theory \cite{t1993planar} at strong coupling, the QCD scalar glueball mass spectra    (specifically the non-supersymmetric Yang-Mills gauge theory) in three and four dimensions were calculated holographically 
\cite{Csaki:1998qr}. This was done by solving the scalar Kaluza-Klein mode equation in the soliton geometry derived from   $\mathrm{AdS}_5 \times \mathbf{S}^5$, where   type IIB superstring theory resides, in the supergravity approximation and from $\operatorname{AdS}_7 \times \mathbf{S}^4$, where   M-theory resides. The obtained mass ratio is in agreement with those in the lattice gauge theory in the continuum limit. The glueball masses of other supergravity models were also investigated  \cite{Minahan:1998tm}. 
In other words, the normal mode frequencies that are  related to the glueball mass were studied via 
Wentzel–Kramers–Brillouin 
(WKB) methods \cite{Schutz:1985km} for massless scalar field perturbations  \cite{Csaki:1998qr,Minahan:1998tm} and for spin-two gravitational perturbations \cite{Constable:1999gb,Brower:1999nj,Garbiso:2020dys}.   The  AdS soliton was also used \cite{Myers:2017sxr} as a holographic model to study the far-from-equilibrium behaviour of the confined system such as the one-dimensional spin chain \cite{kormos2017real}. It has been discovered that for  weak quenches, the  spectrum of normal modes of a massless scalar field for the soliton can be recovered from the response functions. One other intriguing study is a dynamic stability analysis \cite{Durgut:2022xzw} of the Eguchi-Hanson-AdS
soliton spacetime \cite{Clarkson:2006zk,Clarkson:2005qx}.

The soliton obtained from the planar Schwarzschild AdS black hole by analytic continuation can be either in Euclidean or Lorentzian signature. For the Euclidean soliton, a perturbative analysis (for high frequency modes using WKB methods) has been implemented in \cite{Csaki:1998qr,Minahan:1998tm,Constable:1999gb,Brower:1999nj,Garbiso:2020dys}, aiming to study the mass spectrum of the glueball in the dual field theory. For Lorentzian soliton, perturbative analyses in \cite{Nishioka:2009zj,Cai:2011qm,Cai:2011tm} have primarily focused on construction of holographic insulator/superconductor phase transition by finding the zero-frequency quasi-normal mode. Perturbation study for the direct purpose of testing the dynamical stability of the soliton, especially for the general charged soliton \cite{Anabalon:2021tua},  has been less common.  

In this paper, we aim to fill this gap. Recent thermodynamic studies of solitons have been extended to the charged case \cite{Anabalon:2022ksf,Quijada:2023fkc}, which calls for a corresponding dynamic analysis. Our investigation will therefore be grounded in a more general setting, analyzing  massive charged scalar perturbations of the charged AdS soliton, which can be obtained from the planar charged Schwarzschild AdS black hole via analytic continuation. For the charged soliton,  there is a magnetic flux through the compact circle with fixed periodicity in the bulk and a Wilson line with fixed size in the boundary CFT, which is related to the bulk  magnetic flux \cite{Anabalon:2021tua}. The study of the normal modes for the soliton, which are small fluctuations, is a direct test of its dynamic  stability. It can be used to determine whether the  energy of the soliton is a local minimum, thus providing  evidence relevant to the positive energy conjecture \cite{Horowitz:1998ha}.

Our paper is organized as follows. In section \ref{jfeq938pj4}, we review the charged AdS soliton in $D$-dimensional spacetime and formulate the equation of motion for  massive charged scalar perturbations in this background. We next work out the possible boundary conditions for the radial part of the perturbation by local analysis in section \ref{sec:bd}. In section \ref{sec:spacing}, we analyze the relation between light bouncing time and asymptotic modes spacing. In section \ref{jf983p4iou}, we focus on the normal modes of four-dimensional and five-dimensional solitons. The four-dimensional solitons, as solutions to the field equations of Einstein-Maxwell theory, are of purely gravitational interest, and the five-dimensional solitons are relevant in the context of the AdS/CFT correspondence. We find that charged solitons are stable against massive charged scalar perturbations, commensurate with the charged AdS soliton being the minimum energy solution and 
in support of the 
positive energy conjecture. Section \ref{jfopei3j} will be devoted to conclusion and discussions.

\section{Massive charged scalar field perturbation of soliton}\label{jfeq938pj4}

\subsection{Soliton geometry}

The planar charged AdS soliton of interest is described by the metric \cite{Anabalon:2021tua,Yang:2023qxx} 
\begin{equation}\label{solitonmetric} 
\mathrm{d}s^2 = \frac{r^2}{\ell^2}\left(-\mathrm{d}t^2+\sum_{a=1}^{D-3} (\mathrm{d}z^a)^2\right) + \frac{\mathrm{d}r^2}{f(r)} + f(r) \mathrm{d}\phi^2\,, 
\end{equation}
in $D$-dimensional spacetime, 
where $\ell$ is the AdS length scale, $z^a$ are the spatial directions, and $\phi$ is an angular coordinate. The metric function $f(r)$ is given by
\begin{equation}\label{f} 
f(r) = \frac{r^2}{\ell^2} -\frac{\mu}{r^{D-3}} - \frac{Q^2}{r^{2(D-3)}}\,,
\end{equation}
where $\mu$ and $Q$ are two constants. As long as $Q$ and $\mu$ are not simultaneously $0$, by Descartes’s rule of signs, the metric function $f(r)$  has
exactly one positive zero point $r_+$, identified as the position of the soliton surface (or bubble, infrared floor). No spacetime exists
for $r<r_+$, since the signature of the metric would no longer be Lorentzian in that region. There is a conical singularity at $r_+$, which can be removed provided the coordinate $\phi$ is identified as $\phi+\Delta \phi$, with \cite{Horowitz:1998ha,Haehl:2012tw}
\begin{equation}\label{period}
    \Delta \phi = \frac{4\pi}{f'(r_+)} =\frac{2\pi}{\kappa} =\frac{4\pi r_+^{2D-5}\ell^2}{(D-1)r_+^{2D-4}+(D-3)Q^2\ell^2}\,, 
\end{equation}
where $\kappa \equiv f'(r_+)/2$ is   analogous to surface gravity.
With the remaining coordinates being non-compact, the soliton has topology $\mathbf{S}^1\times\mathbb{R}^{D-1}$, and it is locally asymptotically AdS. There is no horizon for this geometry (so its entropy vanishes in the classical limit, as for a field theory in its confined phase), and it can be visualized as one-half of a cigar's surface \cite{Haehl:2012tw} for the $r-\phi$ sector.

The metric \eqref{solitonmetric} solves the equations of motion derived from the $D$-dimensional action for gravity coupled to a Maxwell field
\begin{equation} \label{action}
S=\int \mathrm{d}^Dx\sqrt{-g}\bigg[R-\frac{1}{4}F_{\mu\nu}F^{\mu\nu}+\frac{(D-1)(D-2)}{\ell^2}\bigg]\,. 
\end{equation}
The electromagnetic tensor $F_{\mu\nu}$ has the  gauge potential
\begin{equation}\label{vectorpotential} 
\mathbf{A} = A_\phi \mathrm{d}\phi = \sqrt{\frac{2D-4}{D-3}}\bigg(\frac{Q}{r^{D-3}}-\frac{Q}{r_+^{D-3}} \bigg) \mathrm{d}\phi 
\end{equation}
via $\mathbf{F} = \mathrm{d}\mathbf{A}$. It is obvious that the metric \eqref{solitonmetric} is a saddle point of \eqref{action}, since the whole system is identical to a planar RN-AdS black hole if we apply the swap  
\begin{equation}\label{swp}
    Q\rightarrow iQ\,, \quad t\rightarrow -i\phi\,, \quad \phi \rightarrow it
\end{equation}
to \eqref{solitonmetric} and \eqref{vectorpotential}.
As such, the parameter $Q$ is now interpreted as a magnetic charge, associated with a net magnetic flux $\Phi$ along the $z^a$ axis,
\begin{equation}
    \Phi=-\oint A_{\phi}(r=\infty)\mathrm{d}\phi=\sqrt{\frac{2D-4}{D-3}}\frac{Q}{r_+^{D-3}}\Delta \phi\,.
\end{equation}
Meanwhile, based on the Fefferman–Graham expansion of the metric \eqref{solitonmetric}, the $tt$-component of the holographic stress energy tensor reads
\begin{equation}\label{Ttt}
    \langle T_{tt} \rangle=-\frac{\mu}{2\ell^2}\,,
\end{equation}
indicating a negative soliton mass for a positive $\mu$. Equivalently, 
to obtain a positive soliton mass
requires
$Q\ell^2/r_+^{2D-4}>1$. Conversely, if $\mu>0$ then
\eqref{Ttt} is   interpreted as the Casimir energy in the AdS/CFT correspondence \cite{Horowitz:1998ha,Anabalon:2021tua}.  
We are particularly interested in cases with $D=4$ and $D=5$. Four-dimensional solitons, as electro-vacuum solutions, hold significant gravitational importance, while five-dimensional solitons are relevant to gauge theory within the framework of the AdS/CFT correspondence.

\subsection{Perturbation equation}
We consider a massive charged scalar field perturbation in the soliton background \eqref{solitonmetric}. The Klein-Gordon equation  for the scalar perturbation $\Psi$ is given by 
\begin{equation}\label{KGeqn}
    (\nabla_{\alpha}-iqA_{\alpha})(\nabla^{\alpha}-iqA^{\alpha}) \Psi=M^2 \Psi\,, \quad M^2\geq0
\end{equation}
where $\nabla_\alpha$ is the background covariant derivative, $\mathbf{A}$ is the vector potential given by \eqref{vectorpotential}, and $q$, $M$ are respectively the electric charge and mass of the scalar filed.  The most general solutions to the wave equation \eqref{KGeqn} take the form 
\begin{equation}\label{psiansatz}
    \Psi(t,r,z_a,\phi) = e^{i m\kappa \phi}e^{i\omega t}e^{i k_{a}z^a}R(r)r^{-\frac{D-2}{2}}\,, \quad r\in[r_+,\infty)\,, \quad m\in\mathbb{Z}\,.
\end{equation}
Since from \eqref{period} $\kappa \Delta\phi = 2\pi$, the parameter $m$  must be an integer so that   $\Psi$ is single-valued in $\phi$. 
Meanwhile, its radial part satisfies the equation given by 
\begin{equation}\label{mastereqn}
f \frac{\mathrm{d}}{\mathrm{d}r}\bigg( f\frac{\mathrm{d} R}{\mathrm{d}r}\bigg)=\bigg[(m\kappa-qA_{\phi})^2-\frac{(\omega^2-k_ak^a)\ell^2f}{r^2}+\frac{D-2}{2}\bigg(\frac{(D-4)f^2}{2r^2}+\frac{ff'}{r}\bigg)+M^2f\bigg]R\,.
\end{equation}

If $q,M$ and $Q$ vanish, the equation becomes its counterpart of the AdS planar black hole \cite{Cardoso:2001vs}, with the exchange of quantum numbers $\omega \leftrightarrow i m\kappa$. 
This is expected because the metric is dual to the AdS planar black hole via \eqref{swp}, and the compactification in $\phi$ direction only has the effect of eliminating some eigenvalues of $m$. In this sense, by defining $n$ as the quantum number associated with $r$, we   expect the mode $\omega(n,m,k_a)$ for solitons can be obtained from the quasi-normal modes of black holes by simply exchanging the quantum numbers $\omega $ and $i m\kappa$. However, knowing this doesn't help in practice since we do not have exact expressions for quasi-normal modes of most black holes except simple scenarios like the BTZ case \cite{Chan:1996yk,Chan:1999sc,Cardoso:2001hn}. 

\begin{figure}
    \centering
    \includegraphics[width=5cm]{ 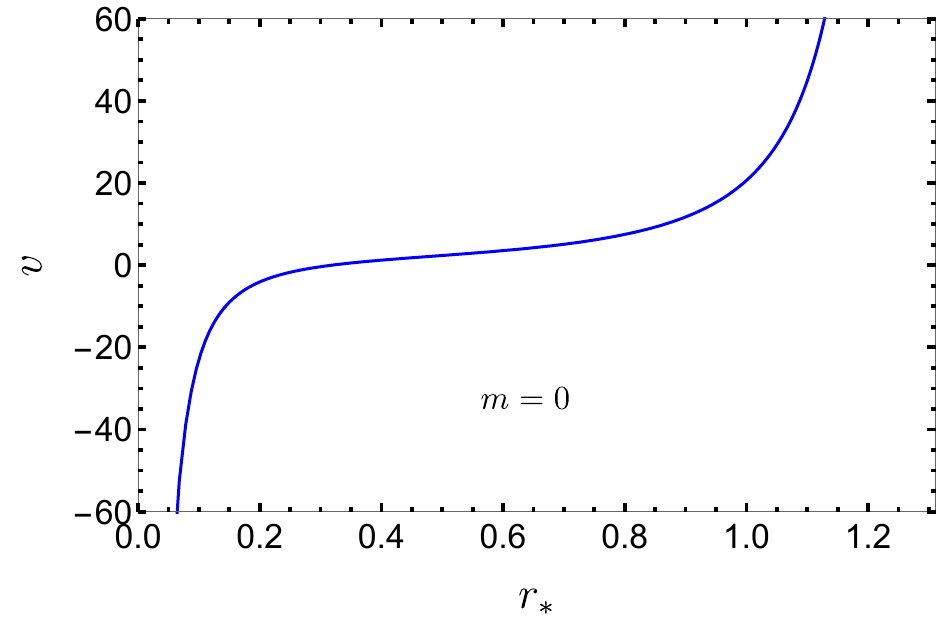}\includegraphics[width=5cm]{ 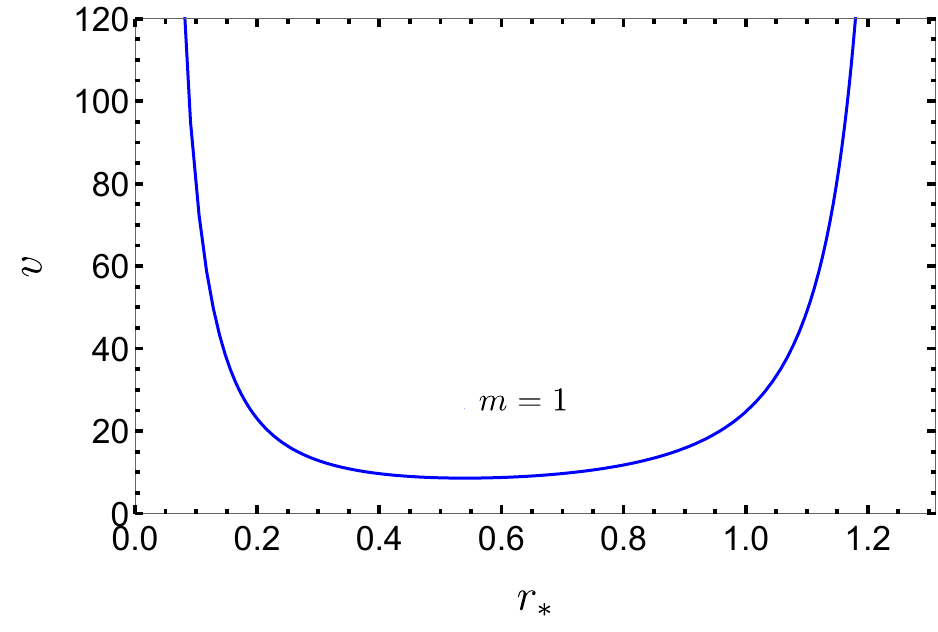}\includegraphics[width=5cm]{ 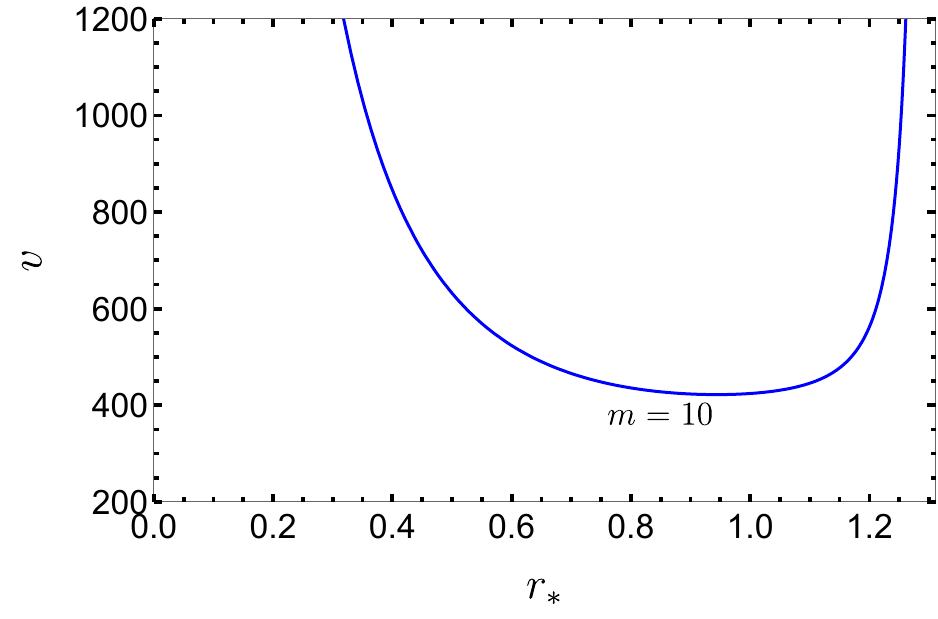}
    \caption{$v$ vs. $r_*$   for different integer $m$'s with $D=4$, $r_+=1$, $Q=1$, $q=0$, $\mu=0$, $\ell=1$. The potential diverges at $0$ and $1.311$. The potential with $m=0$ is the only case where $v$ trends to negative infinity at $r_*=0$. Variations in $D,\, r_+,\, Q,\, q,\,\mu,\, \ell$ do not change the relation qualitatively. }
    \label{fig:v}
\end{figure}

Equation \eqref{mastereqn} is not Schrodinger-like. Hence the second and the third terms in the square bracket no longer play the role of effective potential. To express  \eqref{mastereqn} in the Schrodinger form
\begin{equation}\label{se}
   -\frac{\mathrm{d}^2 \psi}{\mathrm{d}r_*^2} + v(r)\psi = \omega^2\ell^2 \psi\,,
\end{equation}
we apply the coordinate transformation 
\begin{equation}\label{trans}
    r_*(r) \equiv \int_{r_+}^r \frac{\ell \mathrm{d}r'}{r'\sqrt{f(r')}}\,, \quad \psi(r) \equiv r^{-1/2} f^{1/4} R(r)\,,
\end{equation}
and the effective potential $v$ is now identified as 
\begin{equation}\label{effp}
    v(r) \equiv  \frac{r^2(m\kappa -qA_{\phi})^2}{f}-\frac{r^2 f'^2}{16 f}+\frac{1}{4} r^2 f''+\frac{2D-5}{4} r f' + \frac{(D-3)^2f}{4}+k_ak^a\ell^2+M^2r^2\,.
\end{equation}
The tortoise coordinate $r_*$ maps $r$ into a finite region. By the asymptotic integration, it follows that 
\begin{equation}\label{rstar}
    r_*(r) \approx
\begin{cases}
\ell[2r_+^{-2}\kappa^{-1}(r-r_+)]^{1/2}\,, &  r \rightarrow r_+\,; \\
r_*(\infty)-\ell^2/r\,, &   r \rightarrow \infty\,.
\end{cases}
\end{equation}
At these two extremes of $r_*$ the effective potential behaves like 
\begin{equation}\label{v}
    v(r) \sim
\begin{cases}
(4m^2-1)r_+^{3-2D}[(D-3)\ell^2Q^2r_+^4+(D-1)r_+^{2D}]/[16\ell^2(r-r_+)]\,, &r\rightarrow r_+\,; \\ r^2[M^2+D(D-2)/(4\ell^2)]\,,  &r\rightarrow \infty 
\end{cases}
\end{equation}
and we see that $v$ always blows up at both ends, forming two infinite potential walls, as illustrated in figure \ref{fig:v}. 

If $m$ is an integer, 
$\Psi$ satisfies periodic boundary conditions. 
One might expect that a twist ($m=\pm 1/2$) field can have a non-vanishing regular wave  (i.e., $\Psi\sim e^{i \omega (t\pm r_*)}$) at $r_+$, as the potential becomes regular there. However, this is not the case by following the arguments made for boundary conditions in the next section. 

It is noteworthy that massless particles propagating on hyperplanes with constant $z^a$ and $\phi$ are simply characterized by lines $r_*\pm t = \text{constant}$. For AdS \cite{Myers:2017sxr} or Eguchi-Hanson-AdS solitons \cite{Durgut:2022xzw}, the coordinate time $\Delta t=r_*(\infty)-r_*(r_+)$ that a photon needs to reach the AdS boundary from the bubble surface is not only finite but also determines the  gap of high frequency normal modes for massless uncharged particles. In fact, this still holds for the AdS soliton \eqref{solitonmetric} with the magnetic charge generalization, as will be shown in section \ref{sec:spacing}. Moreover, the asymptotic mode spacing is independent of particle charge $q$, the mass $M$ and $m$. 

Before proceeding, we nonetheless set \(k_ak^a = 0\) in the subsequent calculations.
This choice is justified by the fact that, as evident from the factor $(\omega^2-k_ak^a)$ in \eqref{mastereqn}, the physical spectra of \(\omega^2\) for \(k_ak^a \neq 0\) can be straightforwardly obtained by shifting those of \(\omega^2_{k_ak^a=0}\) downward by \(k_ak^a\). 

\section{Local analysis of perturbation equation}\label{sec:bd}
The boundary conditions for $R(r)$ that solve \eqref{mastereqn} can be obtained by a local analysis at $r=r_+$ and $\infty$. To this end, 
we replace $r$ by $x\equiv r_+/r\in[0,1]$, and introduce reduced parameters $\mathfrak{M}=M\ell$, $\mathfrak{Q}=Q\ell r_+^{2-D}$, $\mathfrak{q}=\sqrt{2(D-2)/(D-3)}q\ell$, $\varpi=\omega\ell^2r_+^{-1}$ to get rid of $\ell$ and $r_+$. We thus obtain
\begin{equation}\label{easyform}
    s(x)\hat{R}''(x)+\frac{t(x)}{x -1}\hat{R}'(x)+\frac{u(x)}{(x-1)^2}\hat{R}(x)=0
\end{equation}
where $\hat{R}(x)\equiv R(r(x))$, and coefficient functions $s(x),t(x),u(x)$ are polynomials in $x$, given by 
\begin{equation}
    s(x)=x^2 \mathcal{P}_D^2\,,
\end{equation}
\begin{equation}
    t(x)=x^D\mathcal{P}_D[(D-1)(1-\mathfrak{Q}^2)+2(D-2)\mathfrak{Q}^2 x^{D-3}]\,,
\end{equation}
\begin{equation}
\begin{aligned}
    u(x) = &-\frac{1}{4}x^2 \bigg\{\bigg[(D-3)\mathfrak{Q}^2 +(D-1)\bigg] m+2\mathfrak{q}\mathfrak{Q}(1-x^{D-3})\bigg\}^2\\&+(x-1) \mathcal{P}_D \bigg[-x^2 \varpi^2+\frac{(D-2)(3D-8)}{4}
   \mathfrak{Q}^2  x^{2D-4}\\&\qquad\qquad\qquad\ -\frac{(D-2)^2}{4}(\mathfrak{Q}^2 -1) x^{D-1}+\frac{D(D-2)}{4}+\mathfrak{M}^2\bigg]
\end{aligned}
\end{equation}
with  
\begin{equation}\label{pd}
\begin{aligned}
    \mathcal{P}_D(\mathfrak{Q},x)&\equiv 1+x+x^2+\cdots+ x^{D-2}+\mathfrak{Q}^2 x^{D-1} (1+x+x^2+\cdots+x^{D-4})\\&=
    \frac{x^{D-1}-1}{x-1}+\mathfrak{Q}^2x^{D-1}\frac{x^{D-3}-1}{x-1}\,.
\end{aligned}
\end{equation}
Given $\mathfrak{Q}$, all singularities of \eqref{easyform} are located at the zeros of $\mathcal{P}_D$, together with the points $x=0,1$.  Since $x=0,1$ are regular singularities, the Frobenius method \cite{bender2013advanced} is applicable. The boundary conditions at the end points $x=0,1$ can thus be determined once the corresponding local solutions are obtained.

\subsection{Local solutions at bubble}

At the bubble surface, we have $x=1$. By the Frobenius method, there must exist one solution of the form
\begin{equation}\label{series}
    \hat{R}_1(x)=(x-1)^{\alpha}\sum_{n=0}^{\infty} a_n (x-1)^n, \quad a_0\neq 0
\end{equation}
solving \eqref{easyform} in the vicinity of $x=1$. A non-zero coefficient $a_0$ appears here since we postulate the indicial order $\alpha$ to be the lowest power of $(x-1)$ in the series. Substituting \eqref{series} into \eqref{easyform} and solving \eqref{easyform} order by order yields
\begin{equation}\label{itek}
    a_k=-\frac{1}{P_{kk}}\sum_{n=0}^{k-1} P_{kn}(\alpha)a_n, \quad k\geq 1\,,
\end{equation}
where
\begin{equation}\label{ite1}
 P_{00}(\alpha)=0 
\end{equation}
 and
 \begin{equation}
    P_{kn}(\alpha)\equiv s_{k-n}(n+\alpha)(n+\alpha-1)+t_{k-n}\left(n+\alpha\right)+u_{k-n} 
\end{equation}
for any $k,n\in \mathbb{Z}$, with constants $s_n,t_n,u_n$ individually being Taylor coefficients of $s(x)$, $t(x)$, $u(x)$ expanding around $x=1$. Note that, for any $\alpha$ and integer $k$, we have 
\begin{equation}\label{itid}
    P_{kk}(\alpha)=P_{00}(k+\alpha)\,.
\end{equation}
The exponent $\alpha$ in \eqref{itek} can be obtained by solving the indicial equation \eqref{ite1}, which is one of   
\begin{equation}\label{indicalorder}
    \alpha_1  = \frac{|m|}{2}\,, \quad \alpha_2  = -\frac{|m|}{2}
\end{equation}
with $m$ as in \eqref{psiansatz}.

If the spacing $\Delta \alpha\equiv\alpha_1-\alpha_2$ is a non-integer, a second solution of \eqref{easyform} should also be of the Frobenius form \eqref{series} but with a different indicial order, such as for a twisted scalar field where $|m|=1/4$; the series \eqref{series} with $\alpha=\alpha_2$ is obviously irregular. In our case, $\Delta \alpha=|m|$ is an integer, so the remaining solution could be in other forms. This can be easily seen from the following facts:
if $|m|=0$, namely, $\alpha_1=\alpha_2$, the series \eqref{series} only counts for one independent solution; if $|m|\neq 0$, the denominator $P_{kk}(\alpha_2)$ in \eqref{itek} vanishes for $k=|m|$ since, based on \eqref{itid}, we have 
\begin{equation}
   0 = P_{00}(\alpha_1) =P_{00}(\alpha_2 + |m|) = P_{|m||m|}(\alpha_2)\,,
\end{equation}
which invalidates the iteration \eqref{itek} after $|m|$ steps unless the numerator of $a_{|m|}$ vanishes as well. Therefore the series \eqref{series} with the smaller indicial order $\alpha_2$ generically isn't a solution of \eqref{easyform}. 

We make a few remarks on clarifying what happens if the numerator and denominator of $a_{|m|}$ in \eqref{itek} simultaneously vanish.
It follows that the recursion relation \eqref{itek} still holds for all $k$, but both $a_0$ and $a_{|m|}$ are arbitrary --- the $a_{k<|m|}$ coefficients depend on $a_0$; while the $a_{k>|m|}$ coefficients depend on $a_{|m|}$. This implies the series \eqref{series} contains two independent solutions already, or equivalently, both solutions are in the Frobenius form with indicial orders are precisely given by \eqref{indicalorder}.
 
Aside from this special case, the second solution takes the form 
\begin{equation}\label{series2}
    \hat{R}_2 (x) = \hat{R}_1(x) \ln{(x-1)} + \sum_{n=0}^{\infty} b_n (x-1)^{n+\alpha_2}\,, 
\end{equation}
where $\hat{R}_1$ is the first solution \eqref{series}. As shown in \cite{bender2013advanced}, $b_0$ is absent for $m=0$, but is present for other values of $m$. Therefore, $\hat{R}_2$ is always singular at $x=1$.

In summary, for regular wave modes, the local solution at $x=1$ is of the form \eqref{series}. Therefore we have $\hat{R}(1)=0$ for $m\neq0$ and $\hat{R}(1)=\text{constant}\neq0$ for $m=0$, which means a non-vanishing wave at the edge of the soliton
only appears for $m=0$. 
In the following sections, we focus only on cases with non-negative 
$m$, as it is the absolute value $|m|$ that matters for all regular modes.
 
\subsection{Local solutions at infinity}
At spatial infinity, we have $x=0$. This time we postulate the form of one of the solutions as
\begin{equation}\label{series3}
    \hat{R}_\mathrm{I}(x)=x^{\beta}\sum_{n=0}^{\infty} c_n x^n\,, \quad c_0\neq 0\,. 
\end{equation}
Two indicial orders can be obtained, yielding 
\begin{equation}
    \beta_1 = \frac{1}{2}(1+\sqrt{(D-1)^2+4\mathfrak{M}^2})\,, \quad  \beta_2 = \frac{1}{2}(1-\sqrt{(D-1)^2+4\mathfrak{M}^2})\,.
\end{equation}
The spacing $\Delta \beta\equiv \beta_1-\beta_2=\sqrt{(D-1)^2+4\mathfrak{M}^2}$ implies various possibilities. By the Frobenius theorem again, the series \eqref{series3} with $\beta_1$ always solves   \eqref{easyform}, and the second solution is either of the form
\begin{equation}
    \hat{R}_{\mathrm{II}} = x^{\beta_2}\sum_{n=0}^{\infty} d_n x^n\,, 
\end{equation}
or
\begin{equation}
    \hat{R}_{\mathrm{II}} = \hat{R}_\mathrm{I}\ln{x}+\sum_{n=0}^{\infty} d_n x^{n+\beta_2}\,,
\end{equation}
where $d_0\neq0$. Regularity requires $\hat{R}\sim x^{\beta_{1}}$ at $x=0$, which is equivalent to $\hat{R}(0)=0$. So our wave function must vanish at $r=\infty$. Together with the conclusion drew from last section, we arrive at
\begin{equation}\label{bdconditions}
    \hat{R}(0)=0\,, \quad \hat{R}(1) =\begin{cases}
\text{constant}\neq 0,\quad &m= 0\,, \\
0,\quad &m\neq 0\,.
\end{cases} 
\end{equation}

\subsection{Exact solutions}

According to  Fuchs's theorem \cite{asmar2016partial}, a Frobenius expansion at a point (at least a regular singularity) has a radius of convergence at least  equal to the distance between that point and the nearest singularity in the complex plane.  This implies that the series \( \hat{R}_1 \) could serve as an exact solution of \eqref{easyform} on \( x \in [0,1] \) for certain values of \( \mathfrak{Q}^2 \geq 0 \) that belong to the set  
\begin{equation}
  \Gamma_{D}^1 \equiv \{ \mathfrak{Q}^2 : d_1(\mathfrak{Q}) \geq 1 \wedge \mathfrak{Q}^2 \geq 0 \}\,,  
  \quad d_1(\mathfrak{Q}) \equiv \min \{ |x_i - 1| : \mathcal{P}_D(\mathfrak{Q}, x_i) = 0 \}
\end{equation}  
with \( \mathcal{P}_D \) defined in \eqref{pd}.  It turns out that $\Gamma_{D}^1$ becomes narrower as $D$ increases and finally becomes empty for $D\geq 8$. For cases of our interests, namely, for $D=4,5$, $\Gamma_{4}^1=[0,6]$ and $\Gamma_{5}^1=[0,1]$, which indicate \( \hat{R}_1 \) serves as a global solution when $\mathfrak{Q}^2$ is small enough. 

Similarly, there is also a global solution \( \hat{R}_I \) for $\mathfrak{Q}^2\in \Gamma_{D}^0$, where   
\begin{equation}
\Gamma_{D}^0 \equiv \{ \mathfrak{Q}^2 : d_0(\mathfrak{Q}) \geq 1 \wedge \mathfrak{Q}^2 \geq 0 \}\,, \quad d_0(\gamma) \equiv \min \{ |x_i| : \mathcal{P}_D(\mathfrak{Q},x_i) = 0 \}\,.
\end{equation}  
Interestingly, $\Gamma_{D}^0=[0,1]$ is independent of $D$. Since it is more restrictive for $D=4,5$, we are not particularly interested in \( \hat{R}_I \). Therefore we can always obtain exact solution by the Frobenius method for any soliton with a negative mass since it is equivalent to $\mathfrak{Q}^2\leq1$. 

\section{Light bouncing time and asymptotic modes spacing }\label{sec:spacing}
We reveal intuitively the relationship between the light bouncing time and the asymptotic energetic mode spacing in this section. Owing to the boundary conditions \eqref{bdconditions}, the reduced wave function $\psi$ defined in \eqref{se} must satisfy 
\begin{equation}\label{bd2}
    \psi(r_*(r_+)) = \psi(r_*(\infty)) = 0
\end{equation}
for all $m$. This means that any regular solution $\Psi$ can be interpreted as a bound state $\psi$ that is confined in a compact region $[r_*(r_+),r_*(\infty)]$ with the effective potential $v$ blowing up at both ends, shown in figure \ref{fig:v}. Hence, the system can be identified as a one-dimensional potential well with two hard walls located  at $ r_*(\infty)$ and $r_*(r_+)$. Physically speaking, the eigenvalues $\omega$ should be real, manifesting the conservation of ``energy" conjugate to $t$. 

Recalling that photon travels along the curve where  $r_*\pm t=\mathrm{constant}$,   the spacing of walls  is then simply the coordinate time $\Delta t$ that a photon emitted at $r_+$ takes to reach the AdS boundary, namely
\begin{equation}\label{T}
   \Delta t\equiv r_*(\infty)- r_*(r_+) = \int_{r_+}^{\infty} \frac{ \ell \mathrm{d}r}{r\sqrt{f}} = \int_0^{1}\frac{\ell \mathrm{d}x}{r_+\sqrt{1-\mathfrak{Q}^2x^{2D-4}+(\mathfrak{Q}^2-1)x^{D-1}}}\,.
\end{equation}
For simplicity, we define $\mathcal{T}=r_+\Delta t/\ell$. Thus, in the large $\omega$ limit, the WKB method \cite{Iyer:1986np,Konoplya:2003ii} implies 
\begin{equation}
    n\pi+C=\int_A^B \sqrt{\omega^2\ell^2-v(r)}\mathrm{d}r_* \approx  \int_{r_*(r_+)}^{r_*(\infty)} \omega \ell \mathrm{d}r_* = \varpi\mathcal{T}(\mathfrak{Q})\,,
\end{equation}
where $C$ is some constant, $n$ is the quantum  number associated with $r$ as mentioned above. If $m\neq0$, the two points $A$ and $B(>A)$ are taken to be the turning points where the square root becomes zero; if $m=0$, $A$ is taken to be $r_*(r_+)$ instead. The approximation is taken under two considerations: between the turning points, $v(r)$ is tiny compared to $\omega^2\ell^2$, so an energetic enough  particle feels almost nothing inside the well; when $\omega^2\ell^2$ is large, the turning points get close to $r_*(r_+)$ and $r_*(\infty)$, respectively. Therefore, the asymptotic mode spacing is
\begin{equation}\label{modespace}
    \frac{\Delta \varpi}{\Delta n} = \frac{\pi}{\mathcal{T}}\,, \quad n\rightarrow\infty\,,
\end{equation}
independent of $q$, $M$ and $m$. The explicit values of $\mathcal{T}$ can be evaluated numerically, and decrease as $\mathfrak{Q}^2$ increases. When $\mathfrak{Q}^2 = 0$ or $1$, $\mathcal{T}$ has closed-form expressions  
\begin{equation}
\mathcal{T}(0) = \sqrt{\pi}\frac{\Gamma\left(D/(D-1)\right)}{\Gamma\left(1/2+1/(D-1)\right)}\,, \quad
\mathcal{T}(1) = \frac{\sqrt{\pi}}{3-D}\frac{\Gamma\left(1/(2D-4)\right)}{\Gamma\left(-1/2+1/(2D-4)\right)}\,,
\end{equation}
where $\Gamma$ is the Gamma function.

\section{Numerical solutions of perturbation equation}\label{jf983p4iou}

In this section, two different numerical methods are implemented to evaluate eigenvalues $\omega$ under the boundary conditions in \eqref{bdconditions}, yielding consistent results.

\subsection{Horowitz-Hubeny method}

The Horowitz-Hubeny method \cite{Horowitz:1999jd} can be applied to cases where global solutions are known. By taking $\mathfrak{Q}^2$ from $\Gamma_4^1$ and $\Gamma_5^1$, we obtain $\hat{R}_{1}$ as the solution to  the equation \eqref{easyform} on $x\in [0,1]$ for four and five dimensions, respectively. For these cases, the truncated series can still be a good approximation of $\hat{R}_1$ even at $x=0$. We can thus evaluate $\varpi$ numerically by imposing the boundary condition at $x=0$ to a truncated polynomial of $\hat{R}_1$, and pick out $\varpi$'s that converge as the truncation order  increases.

\subsection{Collocation method}

We use the collocation method for a double check. This method  has no   restrictions on $\mathfrak{Q}^2$ and thus can be applied for any case. To this end, we replace $x$ in \eqref{easyform} by  the transformation 
\begin{equation}
    u\equiv1-2x \in [-1,1]
\end{equation}
for a direct use of standard Chebyshev polynomials. Note that schematically the differential equation \eqref{easyform} takes the form
\begin{equation}\label{eqn}
    \textbf{L} \Tilde{R}(u) = \varpi^2 \Tilde{R}(u)\,, \quad \Tilde{R}(u) \equiv \hat{R}(x)
\end{equation}
with $\textbf{L}$ the corresponding linear differential operator. To incorporate Chebychev polynomials with the boundary conditions \eqref{bdconditions}, we choose 
\begin{equation}
     \phi_j(u)= \begin{cases}
(1-u)(1+u) T_j(u)\,,\quad &m\neq 0\,, \\
(1-u) T_j(u)\,,\quad &m=0
\end{cases} 
\end{equation}
as our basis functions. Solutions $\Tilde{R}(u)$ can thus be approximated by a truncated sum
\begin{equation}\label{apxsol}
    \Tilde{R}(u) \approx \sum_{j=0}^{N-1}c_j \phi_j(u)\,,
\end{equation}
where the coefficients $c_j$ are determined by requiring that \eqref{apxsol} 
 solves  \eqref{eqn} at the collocation points 
\begin{equation}
    u_i = \cos{\left(\frac{i\pi }{N-1}\right)}\,, \quad i = 0,1, \cdots N-1
\end{equation}
on the interval $u \in [-1,1]$. Then we end up with a situation that $\varpi$'s are determined by a generalized algebraic  eigenvalue problem
\begin{equation}
    \textbf{L} \textbf{v} = \varpi^2 \textbf{M} \textbf{v}\,,
\end{equation}
where the matrices are given by
\begin{equation}
    \textbf{L}_{ij}\equiv L[\phi_j(u_i)]\,, \quad \textbf{M}_{ij} \equiv \phi_j(u_i)\,, \quad \textbf{v} \equiv \left[v_0,v_1\,, \cdots, v_{N-1}\right]^{\mathrm{T}}\,.
\end{equation}
In the end, we pick $\varpi$'s that converge as $N$ increases.

\subsection{Normal mode spectra}

The eigenvalues $\varpi$ as normal mode spectra generically   depend on $\mathfrak{Q}$, $\mathfrak{q}$, $\mathfrak{M}$, $m$, $n$. Since the charges $\mathfrak{Q}$ and $\mathfrak{q}$ appear as either $\mathfrak{Q}^2$ or $\mathfrak{Q}\mathfrak{q}$ in the equation of motion \eqref{easyform}, we are free to restrict $\mathfrak{Q}$ to non-negative values and take all real values of $\mathfrak{q}$. Both $\mathfrak{Q}^2$ and $\mathfrak{Q}\mathfrak{q}$ feature the effective potential $v$, but $\mathfrak{Q}^2$ alone determines the well width characterized by $\Delta t$ in \eqref{T}. 
\begin{figure}
    \centering
    \includegraphics[width=7.5cm]{ 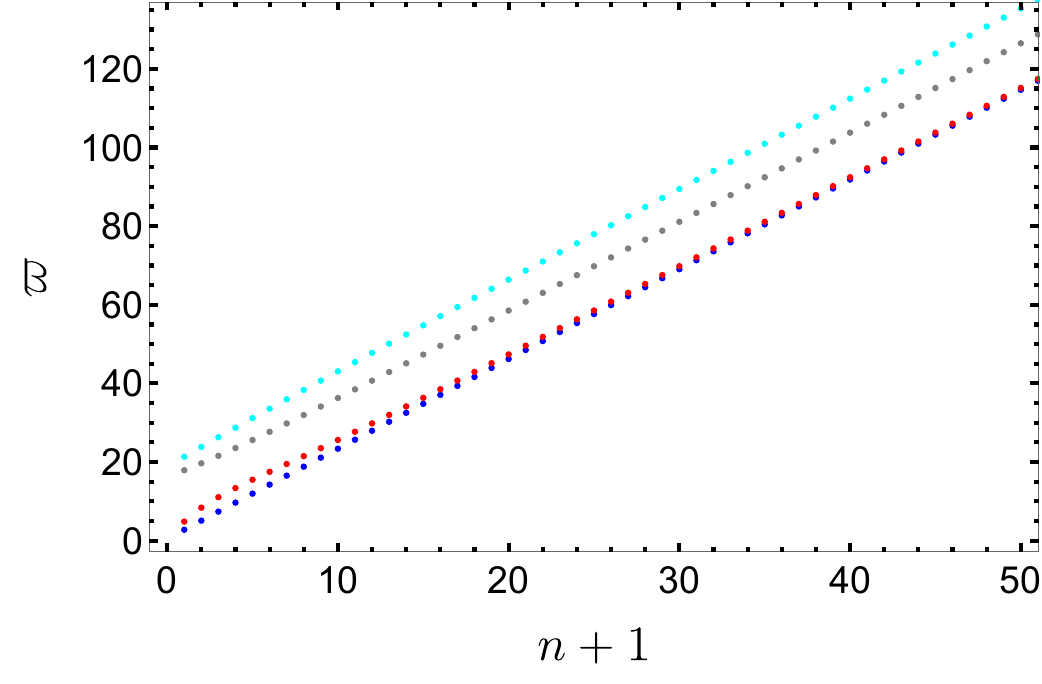}\includegraphics[width=7.5cm]{ 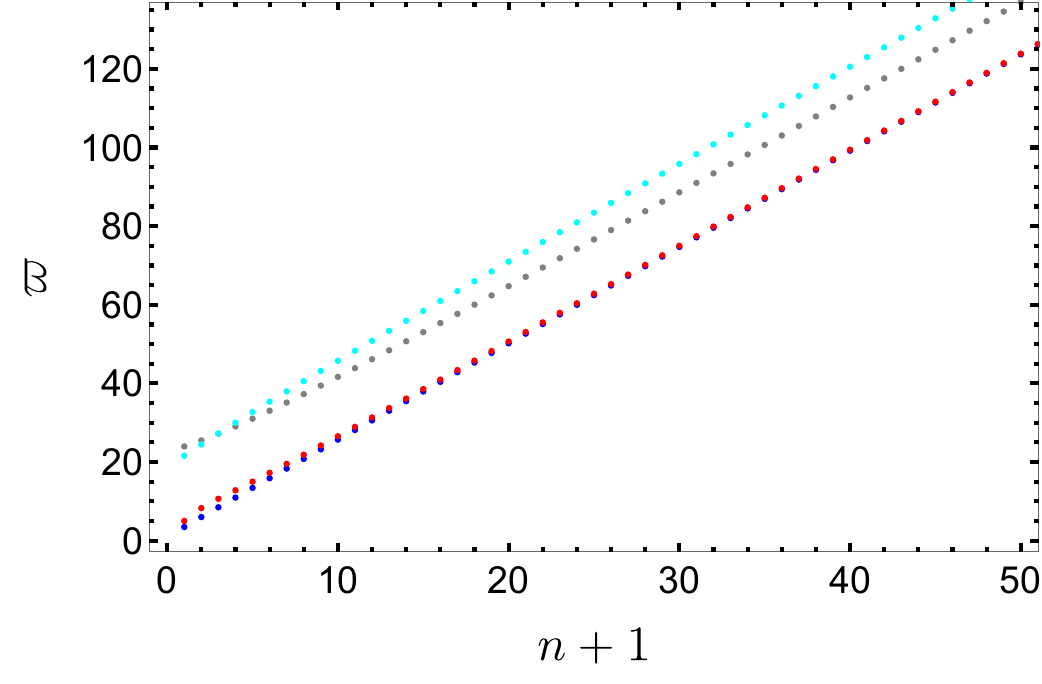}
    \caption{$\varpi$ vs. $n+1$ with $\mathfrak{Q}=1/2$ for $D=4$ (left) and $D=5$ (right). Blue is the basic case with $m=\mathfrak{q}=\mathfrak{M}=0$; the rest have a single change in parameters with respect to the blue points. Red is the same but with $q=-20$, gray is the same but with $m=10$, and cyan is the same but with $\mathfrak{M}=20$. The slope of each line is precisely $\pi/\mathcal{T}(1/2)$, which is  $2.28232$ for $D=4$ and $2.44949$ for $D=5$. Alternations in $\mathfrak{Q}$ don't change these plots qualitatively. }
    \label{fig:omega_n}
\end{figure} 

In figure \ref{fig:omega_n}, we present $\varpi$ as a function of the integer $n$ under various conditions. As $n$ approaches infinity, $\varpi$ becomes asymptotically linear, with its slope precisely determined by \eqref{modespace} and its vertical intercept independent of $\mathfrak{q}$. This behavior is evident from figure \ref{fig:omega_n}, where the blue and red cases exhibit the same asymptotic behavior despite having distinct values of $\mathfrak{q}$. Accordingly, we have
\begin{equation} 
\varpi \sim \frac{n\pi }{\mathcal{T}(\mathfrak{Q})} +\mathcal{C}(\mathfrak{Q}, \mathfrak{M}, m)\,, \quad n\rightarrow +\infty
\end{equation} 
for some function $\mathcal{C}$.

Analogous to free particles in the absence of gravity, where the dispersion relation is given by $\omega = \sqrt{p_{\alpha}p^{\alpha} + M^2}$, we can interpret $n$ as the quantum number associated with the momentum conjugate to $r$. In this context, the first term in the asymptotic dispersion relation aligns with that of relativistic free particles as $n$ becomes large. The second term, $\mathcal{C}(\mathfrak{Q},\mathfrak{M}, m)$, can be understood as a gravitational effect of the soliton that is absent in the Minkowski spacetime.
\begin{figure}
    \centering
    \includegraphics[width=7.5cm]{ 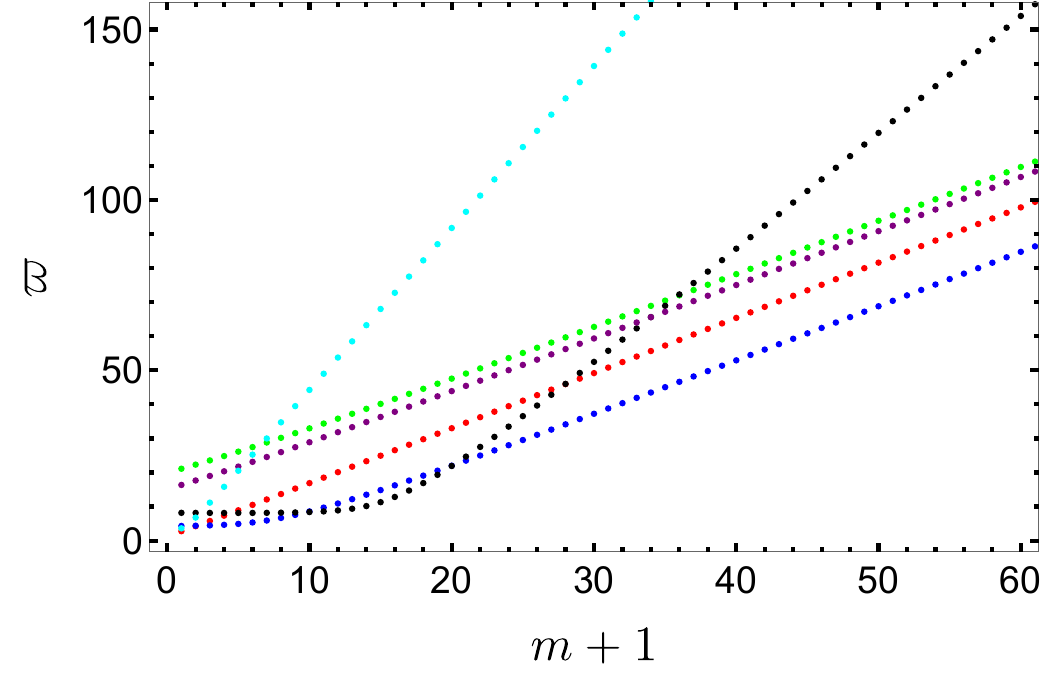}\includegraphics[width=7.5cm]{ 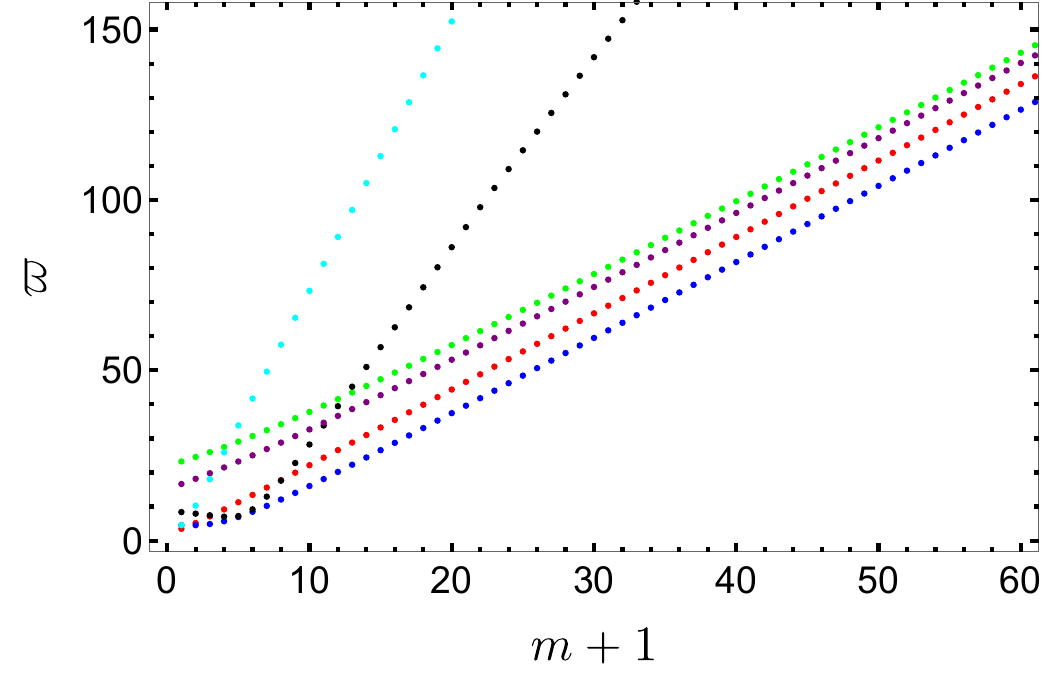}
    \caption{$\varpi$ vs. $m+1$ for $D=4$ (left) and $D=5$ (right). Red is the basic case with $n=\mathfrak{M}=0$, $\mathfrak{Q}=1/2$, $\mathfrak{q}=1$, and  blue, green, and purple have changes in only one parameter relative to red. Blue has $\mathfrak{q}=-15$, green has $n=8$, and purple has $\mathfrak{M}=15$. Cyan and black have $n=\mathfrak{M}=0$ but $\mathfrak{Q}=3, \mathfrak{q}=0$, and  $\mathfrak{Q}=2, \mathfrak{q}=-15$, respectively. The spacing between two successive modes asymptotically approaches a constant depending on $\mathfrak{Q}$ only when $m\to \infty$.}
    \label{fig:oemga_m}
\end{figure}   

\begin{figure}
    \centering
    \includegraphics[width=7.5cm]{ 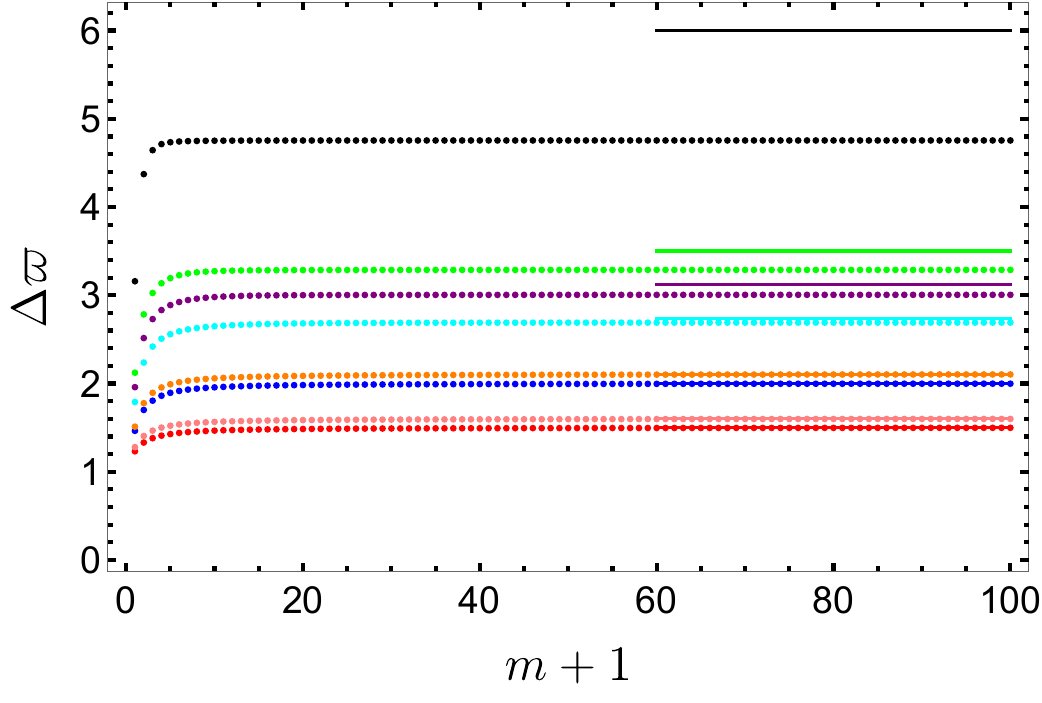}\includegraphics[width=7.5cm]{ 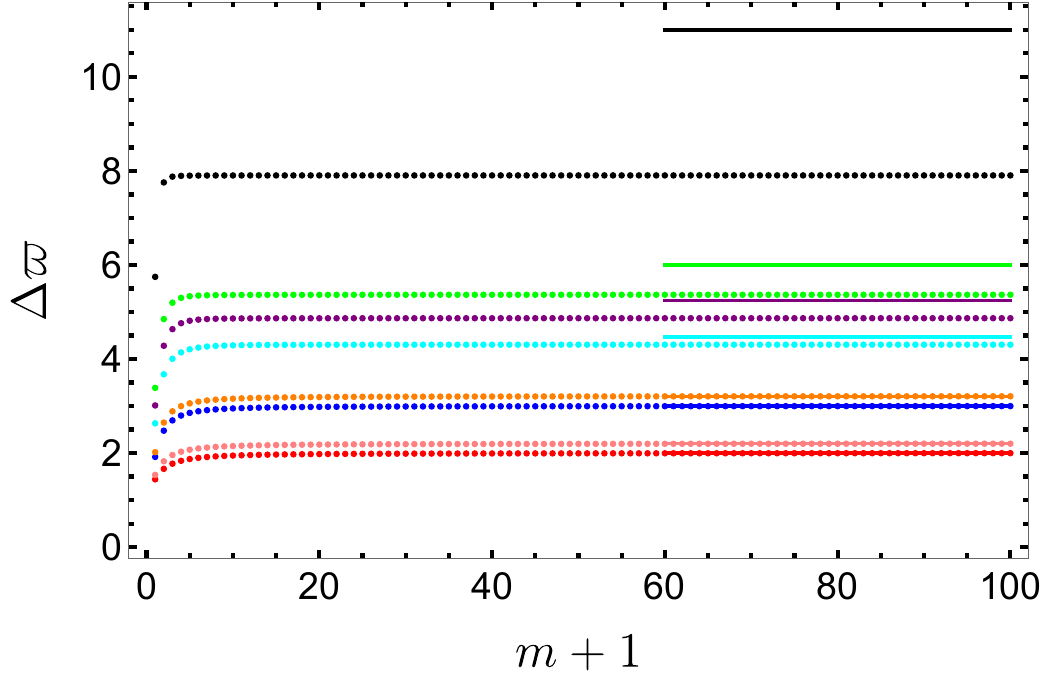}
    \caption{$\Delta\varpi$ vs. $m+1$ for $D=4$ (left) and $D=5$ (right). Dots represent computed values of $\Delta\varpi$, while the solid horizontal lines indicate corresponding values of $\ell^2 \kappa / r_+$; the same color corresponds to the same value of $\mathfrak{Q}$. Since $\Delta\varpi$ depends solely on $\mathfrak{Q}$, all other parameters are set to zero except for $\mathfrak{Q}$. Color codes for $\mathfrak{Q}$ values are: red $=0$, pink $=\pi/7$, blue $=1$, orange $=1.1$;, cyan $=\pi/2$, purple $=1.8$, green $=2$, black $=3$. 
    }
    \label{fig:doemga_m}
\end{figure}   

Similarly, we shall expect $\varpi \propto m$ in the large-$m$ limit since $m$ is related to the conjugate momentum associated with $\phi$. Indeed, as shown in figure \ref{fig:oemga_m}, we observe
\begin{equation}
\varpi \sim \Delta\varpi_m(\mathfrak{Q}) m + \mathcal{B}(\mathfrak{Q},\mathfrak{q}, \mathfrak{M}, n)\,, \quad m \rightarrow +\infty\,,
\end{equation}
where  $\mathcal{B}$ is some subleading term. Notably, the asymptotic proportionality factor $\Delta\varpi_m(\mathfrak{Q})$ depends solely on $\mathfrak{Q}$. It is then natural to guess $\Delta\varpi_m=\ell^2\kappa/r_+$, as it carries the correct unit, as well as depending on $\mathfrak{Q}$ only:
\begin{equation}
    \kappa=\frac{D-1+(D-3)\mathfrak{Q}^2}{2}\frac{r_+}{\ell^2}\,.
\end{equation}
Interestingly, it is not the whole picture. In figure \ref{fig:doemga_m}, we plot the difference between two successive modes $\Delta \varpi=\varpi_{m+1}-\varpi_{m}$ with respect to $m+1$. As the constant lines are $\ell^2\kappa/r_+$, and colors distinguish values of $\mathfrak{Q}$, we see that $\Delta\varpi_m\approx\ell^2\kappa/r_+$ only when $\mathfrak{Q}$ is small. Significant deviation appears when $\mathfrak{Q}\geq3$.  The dependence of $\mathcal{B}$ on $n, \mathfrak{q}, \mathfrak{M}$ is illustrated by the deviations of asymptotes of curves with different $n, \mathfrak{q}, \mathfrak{M}$ in figure \ref{fig:oemga_m}. 

\begin{figure}
    \centering
    \includegraphics[width=7.5cm]{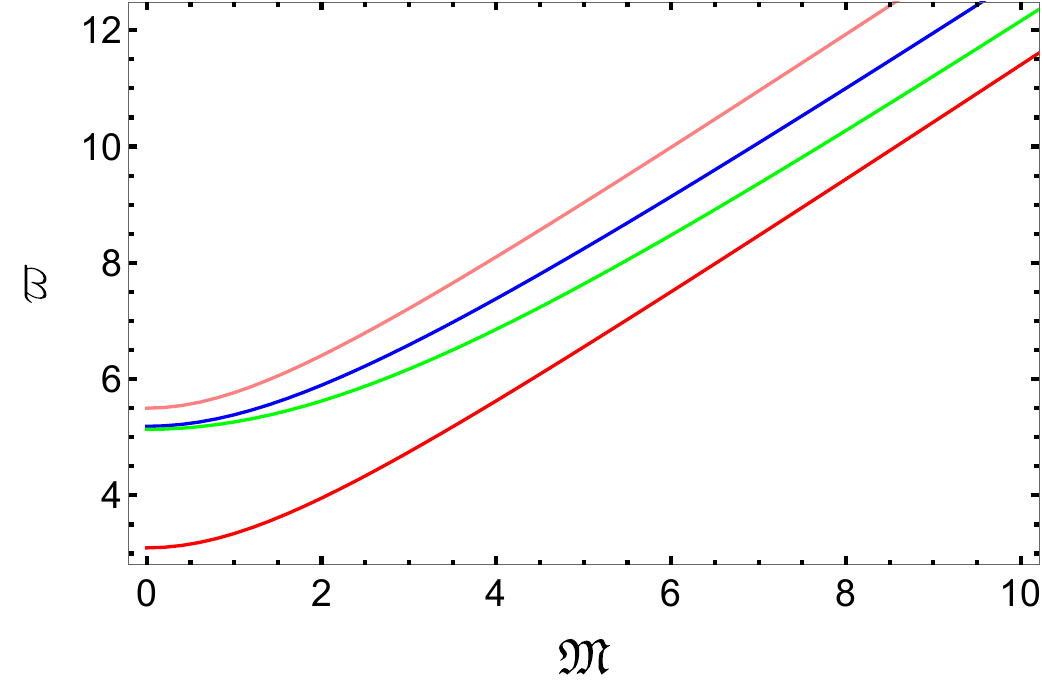}\includegraphics[width=7.5cm]{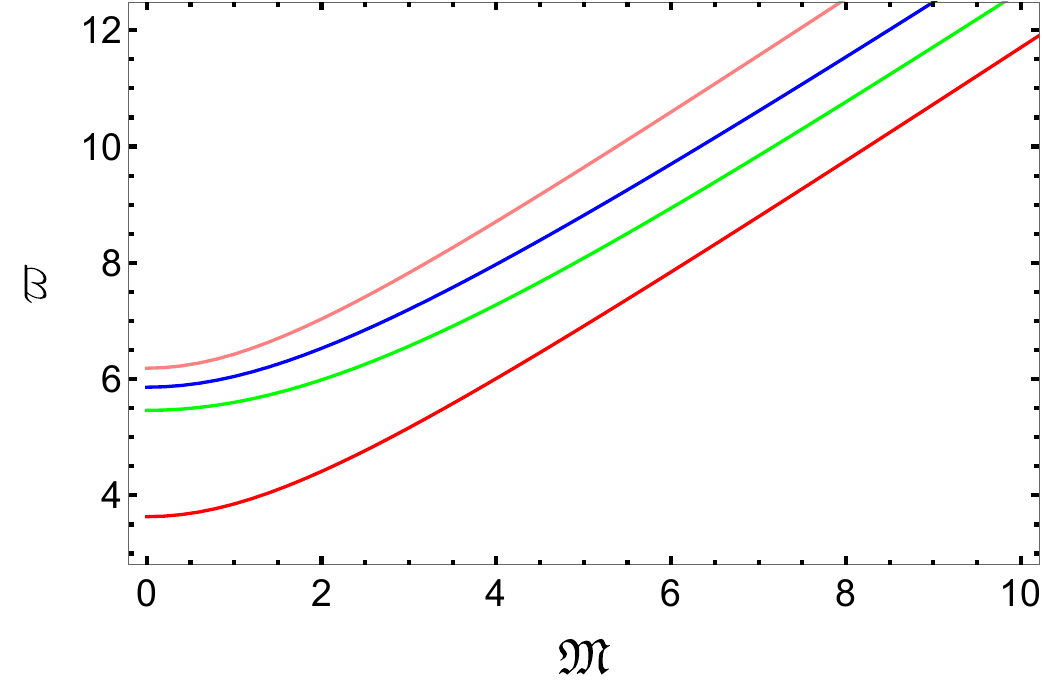}
    \caption{$\varpi$ vs. $\mathfrak{M}$ for $D=4$ (left) and $D=5$ (right) with same $\mathfrak{q}=5$. Red is for $n=m=0$, $\mathfrak{Q}=1/2$, blue is for $n=0$, $m=1$, $\mathfrak{Q}=1/2$, green is for $n=m=0$, $\mathfrak{Q}=2$, pink is for $n=1$, $m=0$, $\mathfrak{Q}=1/2$. The frequency $\varpi$ is proportional to $\mathfrak{M}$ when $\mathfrak{M}$ is large, and the proportionality constant is precisely $1$.}
    \label{fig:oemga_mu}
\end{figure}   

The large-$\mathfrak{M}$ behavior of $\varpi$ is illustrated in figure \ref{fig:oemga_mu}, which reveals an asymptotic dispersion relation of the form
\begin{equation}
\varpi \sim \mathfrak{M} + \mathcal{A}(\mathfrak{Q}, m, n)\,, \quad \mathfrak{M} \rightarrow \infty
\end{equation}
that matches the leading-order behavior of free particles. We observe that $\mathcal{A}$ is independent of $\mathfrak{q}$, as observed from the asymptotic behavior  of curves with different values of $n$, $\mathfrak{q}$, $ \mathfrak{M}$ in figure \ref{fig:oemga_mu}.

\begin{figure}
    \centering
    \includegraphics[width=7.5cm]{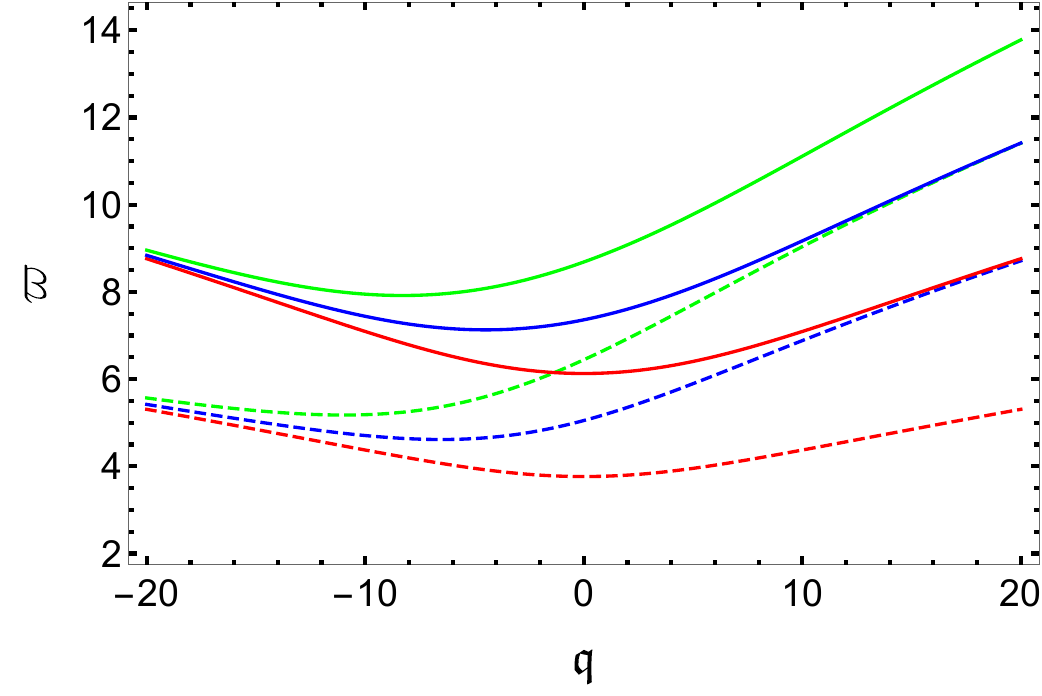}\includegraphics[width=7.5cm]{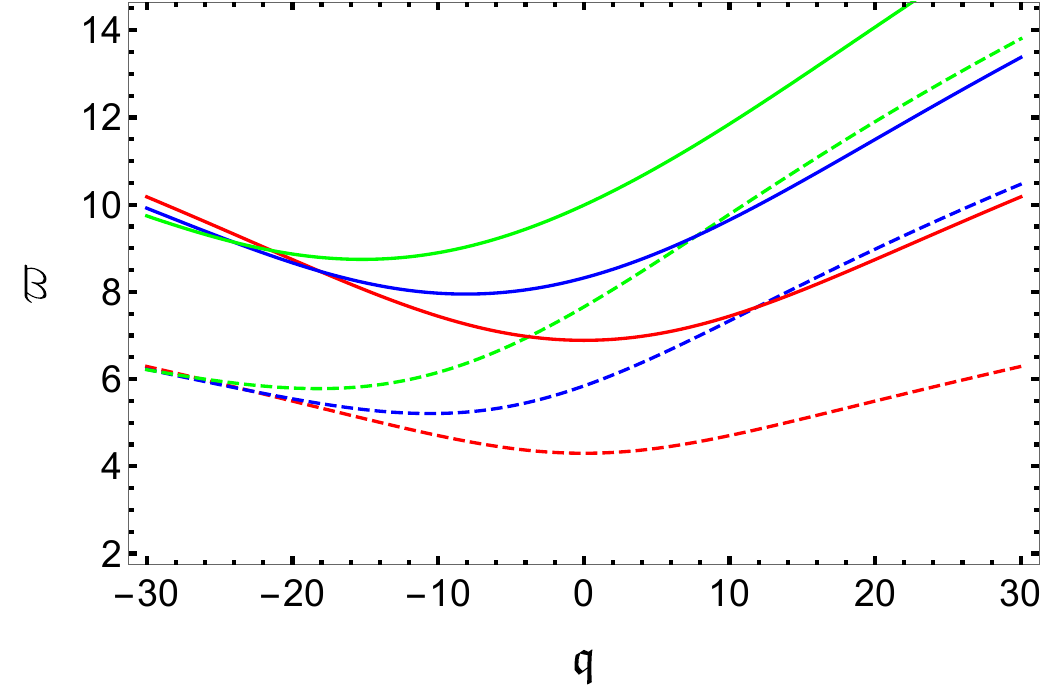}
    \caption{$\varpi$ vs. $\mathfrak{q}$ with $\mathfrak{Q}=1/2$, $\mathfrak{M}=2$ for $D=4$ (left) and $D=5$ (right).  Different colors correspond to distinct  values of $m$: red for $m=0$, blue for $m=1$, and green for $m=2$. Line styles indicate values of $n$: dashed for $n=0$ and solid for $n=1$. }
    \label{fig:oemga_q}
\end{figure}   

Taking $\mathfrak{q}$ to infinity reveals an interesting pattern for $D=4$. As shown in the left graph of figure \ref{fig:oemga_q}, for given values of $\mathfrak{Q}$ and $\mathfrak{M}$, and  fixed $n$, the corresponding values of $\varpi$ exhibit same asymptotic behavior as $\mathfrak{q} \to -\infty$. Additionally, when $m + n$ is held constant, values of $\varpi$ share the same asymptotic behavior as $\mathfrak{q} \to +\infty$. While variations in $\mathfrak{Q}$ and $\mathfrak{M}$ influence the graph quantitatively, they do not alter these overall asymptotic trends. However, this pattern disappears for $D=5$, since in the right graph of figure \ref{fig:oemga_q}, curves intersect as $|\mathfrak{q}|$ gets large. 

\section{Conclusion and discussions}\label{jfopei3j}

We presented a detailed study of the normal modes of charged AdS solitons in $D=4$ and $D=5$, perturbed by massive charged scalar fields. Since soliton geometries differ from black hole solutions only by a coordinate swap, the normal modes of solitons are expected to be dual to the (quasi-normal) modes of black holes, with quantum numbers exchanged according to the swapped coordinates. We verified this duality   rigorously  by comparing the equations of motion for scalar fields in both backgrounds. Using the Horowitz-Hubeny and collocation methods, we numerically investigated how the normal mode spectra rely on the soliton's magnetic charge, the scalar field's mass and charge, and the quantum numbers in both radial and angular directions. Our numerical analysis showed asymptotic behaviors of the normal modes in the limit of large mass, charge and quantum numbers of scalar fields, respectively, showing linear dispersion relations with these numbers, which coincide with the one for a relativistic particle. Notably, in four-dimensional cases, we uncovered a unique asymptotic behavior of the normal modes with respect to the scalar field charge. 

Our results indicate that the charged soliton is stable against massive charged scalar perturbations, showing that it is the lowest energy configuration.  This is commensurate with the positive energy conjecture, which dictates that AdS soliton is the minimum energy solution for an asymptotically locally AdS spacetime  structure.

For future study, one thing is to study the relations between the normal modes of the Lorentzian soliton and the properties of the  Euclidean soliton, which may provides insight to the mass spectrum of the scalar glueball \cite{Constable:1999gb} and other perturbations properties of the dual field theory \cite{Son:2002sd,Birmingham:2001pj}.  The other thing is to study the nonlinear effects of the perturbation to the solitons, as it may offer deeper insights into the dynamic behavior beyond linear approximations, just as studied for black holes \cite{Sberna:2021eui,Cheung:2022rbm,Mitman:2022qdl,Khera:2023oyf,Bucciotti:2023ets,Cardoso:2024jme,Bucciotti:2025rxa,Berti:2025hly}.

\acknowledgments
We are grateful for helpful conversations with Zhen Zhong, Robie Hennigar, Robert Myers, Jiayue Yang, and  Hari Kunduri. This work was supported by the Natural Sciences and Engineering Research Council of Canada, the National Natural Science Foundation of China (Grant No. 12365010), and the China Scholarship Council Scholarship. Research at Perimeter Institute is supported in part by the Government of Canada through the Department of Innovation, Science and Economic Development and by the Province of Ontario through the Ministry of Colleges and Universities.

\bibliographystyle{JHEP}
\bibliography{biblio.bib}

\end{document}